\documentclass[12pt]{article}

\usepackage{amsmath,amssymb,amsthm}
\usepackage[margin=2.5cm]{geometry}
\usepackage{hyperref}
\usepackage{color}
\usepackage{graphicx}

\newcounter{Theorems}
\newcounter{Definitions}
\newcounter{Conjectures}
\begin{document}
\begin{titlepage}
\begin{flushright}

\end{flushright}

\begin{center}
{\Large\bf $ $ \\ $ $ \\
Yang-Mills algebra and symmetry transformations of vertex operators
}\\
\bigskip\bigskip\bigskip
{\large Andrei Mikhailov}
\\
\bigskip\bigskip
{\it Instituto de Fisica Teorica, Universidade Estadual Paulista\\
R. Dr. Bento Teobaldo Ferraz 271, 
Bloco II -- Barra Funda\\
CEP:01140-070 -- Sao Paulo, Brasil\\
}

\vskip 1cm
\end{center}

\begin{abstract}
Linearized solutions of SUGRA equations of motion are
described in the pure spinor formalism by vertex operators.  Under
supersymmetry transformations, they transform covariantly only up to
BRST exact terms.  We identify the cohomology class which is
the obstacle for exact covariance. Computations are simplified by using the
formalism of quadratic algebras.
\end{abstract}

\vfill
{\renewcommand{\arraystretch}{0.8}%
}

\end{titlepage}

\tableofcontents

\section{Introduction}\label{Introduction}

The pure spinor description of the Type II superstring was constructed in \cite{Berkovits:2001ue}.
It has an advantage of having manifest target space supersymmetry in flat spacetime.
This approach may be considered a kind of BRST formalism, but rather unusual one.
The BRST operator is essentially postulated
and does not come in any standard way from gauge fixing applying the Faddeev-Popov procedure.
The ghost fields are constrained;
they take values in a cone, and not in a linear space as in the Faddeev-Popov formalism.
The nonlinear  ghost fields are a characteristic feature of this approach
to superstring theory.

The fields of the corresponding sigma-model are
super-space-time  coordinates $x^m$, $\theta_L^{\alpha}$, $\theta_R^{\hat{\alpha}}$
(where $m$ are coordinate indices of the 10-dimensional spacetime, and
       $\alpha,\hat{\alpha}$ spinor indices) and
the ghosts $\lambda_L^{\alpha}$, $\lambda_R^{\hat{\alpha}}$. The  ``pure spinor'' constraints for ghosts, making them ``nonlinear'', are:
\begin{equation}
     (\lambda_L\Gamma^m\lambda_L) = (\lambda_R\Gamma^m\lambda_R) = 0
     \label{IntroPureSpinorConstraints}\end{equation}
The BRST operator in flat background is:
\begin{equation}
     Q =
     \lambda_L^{\alpha}{\partial\over\partial\theta_L^{\alpha}}
     +
     (\lambda_L\Gamma^m\theta_L){\partial\over\partial x^m}
     +
     \lambda_R^{\hat{\alpha}}{\partial\over\partial\theta_R^{\hat{\alpha}}}
     +
     (\lambda_R\Gamma^m\theta_R){\partial\over\partial x^m}
     \label{IntroBRSTOperator}\end{equation}
Unintegrated vertex operators are expressions of the form:
\begin{equation}
   \lambda_L^{\alpha}\lambda_R^{\hat{\alpha}} A_{\alpha\hat{\beta}}(x,\theta_L,\theta_R)
   \label{IntroVertexOperators}\end{equation}
annihilated by $Q$. They are the 2-cocycles of the BRST complex.
They form a linear space denoted $Z^2$.
This is the space of infinitesimal deformations of the flat background, which is the same as solutions of linearized
SUGRA equations. 
Those which can be represented as $Q W$ for some $W$ are coboundaries.
They correspond to trivial deformations, and  can be removed by gauge transformations.
There is a short exact sequence:
\begin{equation}
   0
   \longrightarrow
   B^2
   \longrightarrow
   Z^2
   \longrightarrow
   [H^2 = \mbox{\tt physical states}]
   \longrightarrow
   0
   \label{IntroShortSequence}\end{equation}
where $B^2$ are coboundaries.
All these spaces are representations of the super-Poincaré group,
and we can consider Eq. (\ref{IntroShortSequence})
is a short exact sequence of representations of the super-Poincaré algebra $\mathfrak{sP}$.
This sequence is not split.
This means that we cannot choose physical vertex operators transforming covariantly
under the super-Poincaré algebra. The covariance only holds up to BRST-exact terms.
There is a cohomological obstacle to covariance of vertex operators, which
was discussed (somewhat implicitly) in \cite{Mikhailov:2011si}. The main purpose
of this letter is to describe this obstacle. 

Let us consider first the case of super-Maxwell theory.
Let $\mathfrak L$ be the 10D Lorentz algebra, $\mathfrak{susy}_1$ the $N=1$ $D=10$ supersymmetry algebra,
and $\mathfrak{sP}_1$ the super-Poincaré algebra. As a linear space $\mathfrak{sP}_1 = \mathfrak{susy}_1 \oplus \mathfrak{L}$.
We denote $\cal O$ the space of local gauge-invariant operators.
The obstacle to constructing covariant vertices was described in \cite{Mikhailov:2011si},
although the supersymmetric completion was not explicitly constructed there. It is:
\begin{equation}
     c^m c^n F_{mn}(0) + (\gamma c^m\Gamma_m \psi(0)) \;\in\; H^2(\mathfrak{sP}, {\cal O}) = (H^2(\mathfrak{susy}_1, {\cal O}))^{\mathfrak{L}}
     \label{IntroObstacleSYM}\end{equation}
Here $F_{mn}=\partial_m A_n - \partial_n A_m$ is the electromagnetic field strength,
and $\psi^{\alpha}$ the fermionic field of the super-Maxwell theory. They are evaluated at a
``marked point'' $0$ in $D=10$ Minkowski space. The cohomology class does not
depend on the choice of the marked point. The upper index $\mathfrak L$ in $(H^2(\mathfrak{susy}_1, {\cal O}))^{\mathfrak{L}}$
means Lorentz-invariant part. The two spaces in Eq. (\ref{IntroObstacleSYM}) coincide because the action of $\mathfrak{L}$
is semisimple.

In the  case of Type II SUGRA,  the obstacle is:
\begin{equation}
     c^m c^n R_{mn\;pq}(0)\otimes t_{[pq]} + \ldots
     \;\in\;
     (H^2\left(\mathfrak{susy}_2, {\cal O}\otimes \mathfrak{sP}\right))^{\mathfrak{L}}
     \label{IntroObstacleSUGRA}\end{equation}
The full expression is rather long, we partially compute it in
Section \ref{Computations}.

Our computation uses an alternative form of the pure spinor BRST complex,
based on the formalism of quadratic algebras.
This approach was introduced in \cite{Chesterman:2002ey}, then used in the super-Yang-Mills
computations in \cite{Movshev:2009ba},
and adapted for SUGRA in \cite{Mikhailov:2012uh}, \cite{Chandia:2013kja} and \cite{Mikhailov:2015sva}.
We feel that it is useful on its own, and perhaps somewhat underappreciated in the pure spinor context.
Our computations here are somewhat in the spirit of \cite{Movshev:2009ba}, but in a simpler context.

We will start by describing the formalism of quadratic algebras.

\section{SYM algebra and SUGRA algebra}\label{IntroAlgebraicApproach}

We consider a quadratic algebra generated by generators $\nabla_{\alpha}$ with a quadratic constraint,
the Koszul dual of the commutative algebra formed by $\lambda^{\alpha}$ with the constraint $(\lambda^{\alpha}\Gamma^m_{\alpha\beta}\lambda^{\beta})=0$.
(The definitions can be found in \cite{Movshev:2009ba},
     and the book \cite{PolishchukPositselski}.)
This algebra has a physical meaning in the large N limit of supersymmetric Yang-Mills theory,
as the algebra formed by covariant derivatives.
A similar algebra, introduced in \cite{Mikhailov:2012uh} and \cite{Mikhailov:2015sva},
is useful in SUGRA computations, for example in this paper. 
Its physical meaning is not very clear. It is \emph{not} the algebra formed by the commutators
of the covariant derivatives of generic curved SUGRA background. 
But formally, it is well-defined, Eqs. (\ref{YangMillsAlgebraLeft}), (\ref{YangMillsAlgebraRight}), (\ref{LeftRightCommute}).

\subsection{SYM algebra}\label{IntroSYMAlgebra}

The SYM algebra is generated by ``letters'' $\nabla_{\alpha}$ subject to the constraint:
\begin{equation}
     \{\nabla_{\alpha},\nabla_{\beta}\} = \Gamma_{\alpha\beta}^m \square_m
     \label{CommutatorConstraint}\end{equation}
This equation is the definition of $\square_m$.
This algebra will be denoted $L$. It has an ideal $I\subset L$ consisting of the elements of the algebra which vanish in the vacuum.
It is generated by $[\square_m,\square_n]$ and $[\nabla_{\alpha},\square_n]$. The factoralgebra is $\mathfrak{susy}_1$:
\begin{equation}
   \mathfrak{susy}_1 = L/I
   \end{equation}
In this language:
\begin{align} F_{mn}(0)\;=\;
 & [\square_m,\square_n]
 \nonumber{} \\ \psi^{\alpha}(0) \;=\;
 & W^{\alpha}\;\;\mbox{\tt where}\quad [\nabla_{\alpha},\square_m] =\Gamma^m_{\alpha\beta}W^{\beta}
 \label{DefW} \end{align}
(Eq. (\ref{DefW}) is the definition of $W$;
     Eq. (\ref{CommutatorConstraint}) implies that the commutator $[\nabla_{\alpha},\square_m]$ is indeed proportional to $\Gamma_{\alpha\beta}^m$, and the coefficient is denoted $W^{\beta}$.)
In fact, the cohomology class in Eq. (\ref{IntroObstacleSYM}) is the obstacle to $L$ being the semi-direct sum
of $I$ and $L/I$. It is not apriori clear why this is also the obstacle to
the existence of a covariant vertex;
we will give explain this coincidence in
Section \ref{sec:ComputationForMaxwell}.

\subsection{SUGRA algebra}\label{IntroSUGRAAlgebra}

The SUGRA algebra is generated by $\nabla_L^{\alpha}$ and $\nabla_R^{\hat{\alpha}}$ with the constraints:
\begin{align}  
 & \{\nabla^L_{\alpha},\nabla^L_{\beta}\} = \Gamma_{\alpha\beta}^m \square^L_m
 \label{YangMillsAlgebraLeft} \\  
 & \{\nabla^R_{\hat{\alpha}},\nabla^R_{\hat{\beta}}\} = \Gamma_{\hat{\alpha}\hat{\beta}}^m \square^R_m
 \label{YangMillsAlgebraRight} \\  
 & \{\nabla^L_{\alpha},\nabla^R_{\hat{\beta}}\} = 0
 \label{LeftRightCommute} \end{align}
In Chapter 3 of \cite{PolishchukPositselski} it is called ``$q$-tensor product'' $\otimes^q$ of two copies
of SYM algebras, for $q=-1$. This is a universal enveloping of a Lie superalgebra which we will denote $\mathfrak{g}$.
It was proven in \cite{PolishchukPositselski}, that the operation $\otimes^q$ preserves the Koszulity property.
This implies that the cohomology of the pure spinor BRST operator  Eq. (\ref{IntroBRSTOperator}) can be
computed as the cohomology of $\mathfrak{g}$ \cite{Mikhailov:2012uh}.

We consider the ideal $\mathfrak{h}\subset\mathfrak{g}$ consisting of those elements which vanish in the vacuum.
This ideal is generated by:
\begin{equation}
   \square^L_m - \square^R_m\;,\;\;
   [\nabla^L_{\alpha},\square^L_m]\;,\;\;
   [\nabla^R_{\hat{\alpha}},\square^R_m]
   \end{equation}
Then $N=2$ SUSY algebra $\mathfrak{susy}_2$ is the factoralgebra:
\begin{equation}
   \mathfrak{susy}_2 = \mathfrak{g}/\mathfrak{h}
   \end{equation}

\section{Pure spinor complex \textit{vs} cohomology of Yang-Mills algebra}\label{Koszul}

The pure spinor complex defined in Eq. (\ref{IntroBRSTOperator}) can be generalized in the following way.
Let $V$ be any representation of the SUGRA algebra. Then we can define the BRST compolex with values in $V$, with the
differential:
\begin{equation}
   Q = \lambda_L^{\alpha}\rho(\nabla_L^{\alpha}) + \lambda_R^{\hat{\alpha}}\rho(\nabla_R^{\hat{\alpha}})
   \end{equation}
where $\rho\;:\; \mathfrak{g} \rightarrow \mbox{End}(V)$ is the action of the representation.
This complex is different from the Lie algebra cohomology complex of $\mathfrak{g}$, but it turns out
that these two complexes are quasi-isomorphic. Given a Lie algebra cocycle, the quasiisomorphism substitutes
$\lambda_L$ and $\lambda_R$ for the ghosts corresponding to $\nabla^L$ and $\nabla^R$, and sets all other ghosts
to zero. The pure spinor constraints, Eq. (\ref{IntroPureSpinorConstraints}) are needed for this to commute with the differential.
It is nontrivial that this map preserves cohomologies, see \cite{Mikhailov:2012uh}, \cite{Mikhailov:2015sva}.
This depends on the special properties of the pure spinor constraint $(\lambda\Gamma^m\lambda)=0$, see [\cite{Gorodentsev:2006fa}].
(Notice that for the nilpotence of the BRST operator of Eq. (\ref{IntroBRSTOperator}), a weaker condition would be needed:
$(\lambda_L\Gamma^m\lambda_L) + (\lambda_R\Gamma^m\lambda_R) = 0$.
But this weaker condition is not the right one;
in particular it would not suffice for this relation to the Lie algebra cohomology, which we need here.)
To summarize:
\begin{equation}
   H^n_{\rm BRST}(V) = H^n(\mathfrak{g},V)
   \end{equation}

\subsection{Lie-algebraic variant of BRST complex}\label{LieAlgebraicBRST}

In particular, in Eq. (\ref{IntroBRSTOperator}) $V$ is the representation of $\mathfrak{g}$ on the space
of functions on Minkowski superspace. We can consider formal Lie group $G$ corresponding to $\mathfrak{g}$.
Similarly, $H$ is the Lie group corresponding to $\mathfrak{h}$:
\begin{equation}
   \mathfrak{g}=\mbox{Lie}(G)\;,\quad    \mathfrak{h}=\mbox{Lie}(H)
   \end{equation}
We work with Taylor series of functions, in the vicinity of some point of spacetime.
We define group elements as exponential of Lie algebra elements, as power series. For our discussion here, we do not have
to worry about the convergence. In other words:
\begin{equation}
   V = C^{\infty}(H\backslash G)
   \end{equation}
Thus, the Lie algebraic version of the BRST complex is:
\begin{equation}
   C^n_{\rm BRST} = C^n\left(\,\mathfrak{g}\,,\,C^{\infty}(H\backslash G)\right)
   \end{equation}
In AKSZ language \cite{Alexandrov:1995kv}:
\begin{align} C^{\bullet}_{\rm BRST} \;=\;
 & C^{\infty}\left(\frac{(H\backslash G)\times \Pi TG}{G}\right)\;\simeq\;
             C^{\infty}(H\backslash \Pi TG)
 \label{LieAlgebraicBRSTComplex} \end{align}
with the differential coming from the canonical vector field $d_{\Pi TG} \in \mbox{Vect}(\Pi TG)$.
We will call this differential $Q_{\rm BRST}$.

We must stress that the BRST complex defined in Eq. (\ref{LieAlgebraicBRSTComplex})
is different from the usual description of the BRST complex, which uses pure spinors.
But there exists a quasiisomorphism between them.

\subsection{Effective Complex}\label{sec:IntroEffectiveComplex}

The cohomology corresponds to the solution of the SUGRA wave equations.
This becomes manifest after ``integrating out'' some BRST-trivial degress of freedom.
More precisely, there is a quasiisomorphism,
see Section \ref{ShapiroLemma} :
\begin{equation}
     i^*\;:\;C^{\infty}\left({\Pi TG\over H}\right) \longrightarrow C^{\infty}\left({\Pi TH\over H}\right)
     \label{PullbackOfRestriction}\end{equation}
--- the $\Pi T(G/H)$ gets ``integrated out''. This is just the restriction to $\Pi TH\subset \Pi TG$.
After that, the wave equations become manifest, see \cite{Mikhailov:2012uh}.

\subsection{Symmetry bicomplex and spectral sequence}\label{sec:SymmetryBicomplex}

Let $\cal O$ denote the space of local gauge invariant operators
at a marked point $0$ of the Minkowski superspace. 

This space is dual to the space of linearized solutions of SUGRA.
Indeed, a linearized solution can be completely characterized, modulo gauge
transformations, evaluating local operators at $0$.

Both $\cal O$ and $C^{\bullet}_{\rm BRST}$ are representations of $\mathfrak{g/h}$.
Consider the complex:
\begin{equation}
     C^{\bullet}\left(\;\mathfrak{g/h}\;,\;{\cal O}\otimes C^{\bullet}_{\rm BRST}\;\right)
     \label{StraightForwardComplex}\end{equation}
with the differential:
\begin{equation}
     d_{\rm tot} = d_{\mathfrak{g/h}} + Q_{\rm BRST}
     \label{StraightForwardDifferential}\end{equation}
Notice that $d_{\mathfrak{g/h}}$ anticommutes with $Q_{\rm BRST}$.
Consider the filtration corresponding to the $\mathfrak{g/h}$-ghost number
(\textit{i.e.} the Lie cohomology degree).
There is the corresponding spectral sequence which we call $E^{p,q}$. In particular:
\begin{equation}
   E_2^{0,2} = \left({\cal O}\otimes H^2_{\rm BRST}\right)^{\mathfrak{g/h}}
   \end{equation}
where $(\ldots)^{\mathfrak{g/h}}$ means $\mathfrak{g/h}$-invariants.
Since $H^2(Q_{\rm BRST})$ is the space of linearized solutions, and $\cal O$ its dual,
${\cal O}\otimes H^2_{\rm BRST}$ can be identified with the space of linear maps
acting in the space of linearized solutions.
(More precisely, linear maps acting on Taylor series expansion of solutions around the marked point.)

\subsection{Obstacle to exact covariance}\label{sec:ObstacleToExactCovariance}

Consider the identity map ${\bf 1}\in E_2^{0,2}$, and its $d_2$:
\begin{align} \Psi \;=\;
 & d_2\,({\bf 1}\;:\;H^2_{\rm BRST}\rightarrow H^2_{\rm BRST})
 \label{ObstacleAsD2Id} \\ \Psi \;\in\;
 & E_2^{2,1} = H^2\left(\,\mathfrak{g}/\mathfrak{h}\,,\,{\cal O}\otimes H^1_{\rm BRST}\,\right)
 \nonumber{} \end{align}
This $\Psi$ is the obstacle to the existence of the covariant vertex. The proof is the same as in [\cite{Mikhailov:2011si}],
except that now we are using the Serre-Hochschild resolution to compute the cohomology of $\mathfrak{g}$.
While in [\cite{Mikhailov:2011si}] the Koszul (pure spinor) resolution was used.

If $d_2{\bf 1}$ were zero, then the obstacle would be $d_3{\bf 1}$,
an element of $H^3\left(\;\mathfrak{sP}\;,\;{\cal O}\;\right)$.
But this is not the case. In fact, $H^3\left(\;\mathfrak{sP}\;,\;{\cal O}\;\right) = 0$.
An obvious guess $H^{NSNS}_{klm}(0)c^kc^lc^m + \ldots$, a naive generalization of Eq. (\ref{IntroObstacleSYM}),
does not have supersymmetric completion (the ``$\ldots$'' does not exist).

There is another explanation for why the obstacle lives in $H^2\left(\,\mathfrak{g}/\mathfrak{h}\,,\,{\cal O}\otimes H^1_{\rm BRST}\,\right)$. It was shown in \cite{Mikhailov:2021jdw} that
the linearized solutions are in one-to-one correspondence with nontrivial infinitesimal deformations
of the nilpotent vector field defined by Eq. (\ref{IntroBRSTOperator}):
\begin{equation}
   Q\;\mapsto\; Q + \varepsilon q
   \end{equation}
Therefore we can consider the identity map:
\begin{equation}
   {\bf 1}\in H^0(\mathfrak{g}/\mathfrak{h},{\cal O}\otimes H^1_{\rm BRST}(\mbox{Vect}(x,\theta,\lambda)))
   \end{equation}
The $d_2$ of it gives us an element of:
\begin{equation}
   H^2(\mathfrak{g}/\mathfrak{h},{\cal O}\otimes H^0_{\rm BRST}(\mbox{Vect}(x,\theta,\lambda)))
   \end{equation}
Here $H^0_{\rm BRST}(\mbox{Vect}(x,\theta,\lambda))$ is the ghost number zero cohomology in the space of vector fields.
It corresponds to symmetries, see \cite{Mikhailov:2021jdw}. It is the same as $H^1_{\rm BRST}(\mbox{Fun}(x,\theta,\lambda))$
which we call just $H^1_{\rm BRST}$.
Here we denote $\mbox{Fun}(x,\theta,\lambda)$ the space of smooth functions of $x,\theta,\lambda$
with polynomial dependence on $\lambda$.

\section{Shapiro's lemma}\label{ShapiroLemma}

It turns out that the complex (\ref{StraightForwardComplex}), (\ref{StraightForwardDifferential}) can be simplified, using the technique
known as ``Shapiro's lemma''. In fact, it is quasiisomorphic to $C^{\bullet}(\mathfrak{g},{\cal O})$.

\subsection{BRST formulation}\label{BRSTformulation}

Using the AKSZ \cite{Alexandrov:1995kv} description, the Serre-Hochschield cochain complex
for $H^m\left(\mathfrak{g},\mbox{Coind}_{\mathfrak h}^{\mathfrak g} \mathbb{C}\right)$
can be identified with functions, more precisely Taylor series near the unit:
\begin{equation}
   C^{\bullet}\left(\mathfrak{g},\mbox{Coind}_{\mathfrak h}^{\mathfrak g} W\right)
   \;=\;
   C^{\infty}(H\backslash G)\otimes_G C^{\infty}(\Pi TG) \simeq C^{\infty}(H\backslash \Pi TG)
   \end{equation}
with the differential induced from the de Rham differential of $\Pi TG$. In this language, the Shapiro's lemma is:
\begin{equation}
     H\left(\;C^{\infty}(H\backslash \Pi TG)\;,\; d_{\Pi TG}\;\right)
     =
     H\left(\;C^{\infty}(H\backslash \Pi TH)\;,\; d_{\Pi TH}\;\right)
   \label{ShapiroBRST}\end{equation}
To understand this, it is useful to think of $C^{\infty}(H\backslash \Pi TG)$ as
the space of differential forms on $G$ left-invariant under $H$.
More precisely, we consider Taylor series of forms in the vicinity of the unity of $G$.
We always work in the vicinity of the unity. Then, we can choose a section of the fiber bundle
$G\rightarrow G/H$. Let us call it $s$:
\begin{equation}
     s\;:\;G/H\rightarrow G
     \label{SectionS}\end{equation}
Such a section defines a map:
\begin{align} \chi\;:\;
 & G\rightarrow H
 \nonumber{} \\ \chi(g)\;=\;
 & gs^{-1}([g])
 \nonumber{} \end{align}
It satisfies, for all $h\in H$:
\begin{equation}
   \chi(hg) = h\chi(g)
   \end{equation}
This defines a linear map (``the pullback''):
\begin{align} \chi^*\;:\;
 & C^{\infty}(H\backslash \Pi TH)\rightarrow C^{\infty}(H\backslash \Pi TG)
 \nonumber{} \end{align}
which commutes with the differential.
This Shapiro's lemma implies that this map is a quasi-isomorphism.
Indeed, the map $i^*$ of Eq. (\ref{PullbackOfRestriction}) satisfies:
\begin{align}  
 & i^*\chi^* = \mbox{id}
 \nonumber{} \\  
 & \chi^* i^* = \mbox{id} + [d,{\cal L}_E^{-1}\iota_E]
 \nonumber{} \end{align}
where $E$ is an $H$-invariant vector field on $G$ which is tangent to the section $s$ and
with restriction on $s$ being an Euler vector field in the vicinity of the unit of $H\backslash G$.
(This is essentially the Poincare lemma in the directions transverse to the orbits of the left action of $H$ on $G$.)

\section{Filtered complex from the action of global symmetries}\label{FilteredComplex}

The SUSY group $G/H$ acts on the BRST complex $C^{\infty}(H\backslash G)\otimes_G C^{\infty}(\Pi TG) \simeq C^{\infty}(H\backslash \Pi TG)$.
In the language of Lie algebra cohomology, there is  the correspondins cochain complex:
\begin{equation}
     C^{\bullet}\left(\;\mathfrak{g}/\mathfrak{h}\;,\;C^{\infty}(H\backslash \Pi TG)\;\right)
     \label{FilteredComplexSymmetries}\end{equation}
Let us define a filtration on this complex, in the following way.
For an element $g\in G$, we write $g = \exp(x^i t_i + y^m t'_m)$ where $t'_m$ run over a basis of $\mathfrak h$
and $\{t_i\}$ complement them to a basis of $\mathfrak g$.
(In other words, $\{t_i \;\mbox{mod}\;\mathfrak{h}\}$ form a basis of $\mathfrak{g}/\mathfrak{h}$.)
We then say that the coordinates $x^i$ have grade $1$, and coordinates $y^m$ have grade $0$.

The cohomology of $C^{\infty}(H\backslash \Pi TG)$ corresponds to solutions of linearized SUGRA equations.
This is seen \cite{Mikhailov:2012uh} by considering the embedding:
\begin{equation}
   H\backslash \Pi TH \rightarrow H\backslash \Pi TG
   \end{equation}
which induces a quasiisomorphism on functions. 
We may think of this as ``integrating out'' the coordinates in $H\backslash \Pi TG$ which are transverse to
$H\backslash \Pi TH$. Then the action of $G/H$ becomes an $L_{\infty}$-action,
Eq. (\ref{CohomologyCohomology}).

The AKSZ description of the  complex (\ref{FilteredComplexSymmetries}) is:
\begin{equation}
   \frac{\Pi T(H\backslash G)\times (H\backslash \Pi TG)}{H\backslash G}
   \simeq
   {\Pi T(H\backslash G)\times \Pi TG \over G}
   \end{equation}
For any supermanifold $X$, the odd tangent space can be thought of as the space of maps $\mathbb{R}^{0|1}\rightarrow X$
(more precisely, using functor of points, we define $\Pi TX$ so that for any supermanifold $Y$:
      $\mbox{Map}(Y, \Pi TX) = \mbox{Map}(Y\times \mathbb{R}^{0|1}, \,X)$).
Let $\zeta$ denote a coordinate on $\mathbb{R}^{0|1}$ (it is Grassmann odd).
Let a $H\backslash G$-valued function $f(\zeta)$ describe a point in $\Pi T(H\backslash G)$, and $g(\zeta)$ a point in $\Pi TG$.
Since $\Pi TG$ acts on $\Pi T(H\backslash G)$, we can construct a map:
\begin{align}  
 & {\Pi T(H\backslash G)\times \Pi TG \over G} \longrightarrow \Pi T(H\backslash G)\times (G\backslash\Pi TG)
 \nonumber{} \\  
 & (\;f(\zeta)\;,\;g(\zeta)\;) \;\mapsto\; (\;g(\zeta)^{-1}f(\zeta)\;,\;g(\zeta)\;)
 \nonumber{} \end{align}
Since $\Pi T(H\backslash G)$ is acyclic, injection of functions constant along $\Pi T(H\backslash G)$ is a quasiisomorphism:
\begin{equation}
     C^{\infty}\left(G\backslash\Pi TG\right)
     \longrightarrow
     C^{\infty}\left(\Pi T(H\backslash G)\times (G\backslash\Pi TG)\right)
     \label{Quasiisomorphism}\end{equation}
Moreover, this is quasiisomorphism on the first page of the spectral sequence.
Indeed, the differential on $\mbox{gr}\,C^{\infty}(\Pi T(H\backslash G))$ becomes the de Rham differential on $\mathfrak{g}/\mathfrak{h}$
(a linear space),  acyclic by the Poincare lemma. On both sides, the first page is
$E_1^{p,q} = P^p\left(\Pi (\mathfrak{g}/\mathfrak{h})\right)\otimes H^q(\mathfrak{h})$,
where we denote $P^p$ the space of polynomials of degree $p$.
Therefore, by the Mapping Lemma 5.2.4 of [\cite{WeibelIntro}], it is also an isomorphism on the second page.

We can generalize by considering the tensor product with any representation of $\mathfrak{g}/\mathfrak{h}$.
In particular, consider the representation in the space of local operators, which we denote $\cal O$.
The map:
\begin{equation}
   C^{\infty}\left(\Pi TG\right)\otimes_G {\cal O} 
   \rightarrow 
   C^{\infty}\left(\Pi T(H\backslash G)\times\Pi TG\right)\otimes_G {\cal O}
   \end{equation}
induces a quasiisomorphism  of the spectral sequence starting from the first page.

This implies that the obstacle Eq. (\ref{ObstacleAsD2Id}) can be computed using the spectral sequence of
the filtered complex $C^{\bullet}\left(\mathfrak{g},{\cal O}\right)$.
We will now explicitly describe the differential in this complex.

Let us pick an injective linear map (a ``section'')
\begin{equation}
   s\;:\;{\mathfrak g}/{\mathfrak h} \rightarrow {\mathfrak g}
   \end{equation}
satisfying:
\begin{equation}
   s(x)\; \mbox{mod}\; \mathfrak{h} = x
   \end{equation}
Define a bilinear function $\sigma$ on ${\mathfrak g}/{\mathfrak h}$ with values in $\mathfrak h$:
\begin{equation}
     \sigma(x\wedge y) = [s(x),s(y)] - s([x,y])\;\in\;\mathfrak{h}
     \label{DefSigma}\end{equation}
The cochain complex of $\mathfrak g$ can be presented in the following way:
\begin{align}  
 & C^{\bullet}(\mathfrak{g})
        \;=\;
        C^{\bullet}\left(\,\mathfrak{g}/\mathfrak{h}\,,\, C^{\bullet}(\mathfrak{h})\,\right)
 \label{CC} \\  
 & d'
        \;=\;
        d_{{\mathfrak g}/{\mathfrak h}}\omega \,+\, d_{{\mathfrak h}} \omega \,+\, {\cal L}_{C(\mathfrak{h})}\langle c^as(t_a)\rangle\omega \,+\, \iota_{C({\mathfrak h})}\langle\sigma(c,c)\rangle\omega
 \label{CohomologyCohomology} \end{align}
where $c^a$ are the Faddeev-Popov ghosts of $\mathfrak{g}/\mathfrak{h}$.
The differential $d'$ can be thought of as defining the $L_{\infty}$-action of $\mathfrak{g}/\mathfrak{h}$
on $C^{\bullet}(\mathfrak{h})$, see Appendices of \cite{Movshev:2009ba}.

\section{Mixed complex}\label{PartialDualization}

\subsection{Definition of mixed complex}\label{sec:MixedComplex}

Consider the following geometrical construction. Let $M$ be a Q-manifold, with the
cohomological vector field $Q_M\in\mbox{Vect}(M)_{\bar{1}}$. Let $N$ be another manifold, with
the following structure:
\begin{align}  
 & A\;\in\;(C^{\infty}(M)\otimes \mbox{Vect}(N))_{\bar{1}}
 \label{MixedCohomologicalVectorField} \\  
 & Q_MA + {1\over 2}[A,A] = 0
 \label{GeometricalMC} \end{align}
Then $Q_M + A$ is a cohomological vector field on $M\times N$. We pick a point $n_0\in N$, and consider
Taylor series in a formal neighborhood of $n_0$ in $N$
We can consider two complexes.
One is $C^{\infty}(M\times N)$ with the differential $Q_M + A$. The second is $C^{\infty}(M)\otimes D(N)$
where $D(N)$ is the space of differential operators on $N$ modulo those operators $O$ which have a property:
$(Of)(n_0)=0$ for all $f\in C^{\infty}(N)$.
The space $D(N)$ is dual to $C^{\infty}(N)$ (more precisely, to Taylor series at $n_0$).
The space  $C^{\infty}(M)\otimes D(N)$ has a differential $Q_M + A'$ where:
\begin{align}  
 & A'\;\in\;(C^{\infty}(M)\otimes \mbox{End}(D(N)))_{\bar{1}}
 \nonumber{} \\  
 & \langle A'\phi,a\rangle = (-)^{\bar{\phi}}\langle \phi, Aa\rangle
 \nonumber{} \end{align}
Here $\phi \in C^{\infty}(M)\otimes D(N)$, $a\in C^{\infty}(M\times N)$ and $\langle,\rangle$ the natural
pairing taking values in $C^{\infty}(M)$.

To apply this to the complex (\ref{CC}), we take $M = \Pi (\mathfrak{g}/\mathfrak{h})$ and $N = \Pi\mathfrak{h}$.
Then $C^{\infty}(M\times N)$ becomes $C^{p,q} = C^p(\mathfrak{g}/\mathfrak{h},C^q(\mathfrak{h}))$, and $C^{\infty}(M)\otimes D(N)$ becomes:
\begin{align} \widetilde{C}^{p,-q} \;=\;
 & C^p(\mathfrak{g}/\mathfrak{h},C_q(\mathfrak{h}))
 \nonumber{} \\ \widetilde{d} \;=\;
 & d_{{\mathfrak g}/{\mathfrak h}}\omega \,+\, \delta_{{\mathfrak h}} \omega \,+\, {\cal L}_{C(\mathfrak{h})}\langle c^as(t_a)\rangle\omega \,+\, \sigma(c,c)\wedge\omega
 \nonumber{} \end{align}
where $C_q(\mathfrak{h}) = \Lambda^q\mathfrak{h}$ and $\delta_{\mathfrak h}$ is the homological differential.
For this construction it is important that ${\mathfrak h}\subset{\mathfrak g}$ is an ideal.

We need a slight generalization of this construction. Let $V$ be a representation of ${\mathfrak g}/{\mathfrak h}$.
Then, consider the complex:
\begin{equation}
   S^{\bullet}\left(\Pi ({\mathfrak g}/{\mathfrak h})^*\right)\otimes C_{\bullet}({\mathfrak h})\otimes V
   \end{equation}
with the differential:
\begin{equation}
     \widetilde{d}\omega = \delta \omega + d_{{\mathfrak g}/{\mathfrak h}}\omega + {\cal L}_{C^as(t_a)}\omega + \rho_V(C^at_a)\omega + \sigma(C,C)\wedge\omega
     \label{MixedDifferential}\end{equation}
There is a filtration by the ghost number $C^a{\partial\over\partial C^a}$. Consider the corresponding spectral sequence $\widetilde{E}_r^{p,-q}$.
The second page is:
\begin{equation}
   \widetilde{E}_2^{p,-q} = H^p(d_{{\mathfrak g}/{\mathfrak h}}, H_q(\delta_{\mathfrak h})\otimes V)
   \end{equation}
with the differential:
\begin{equation}
   \widetilde{d}_2\;:\;
   H^p(d_{{\mathfrak g}/{\mathfrak h}}, H_q(\delta_{\mathfrak h})\otimes V)
   \rightarrow
   H^{p+2}(d_{{\mathfrak g}/{\mathfrak h}}, H_{q+1}(\delta_{\mathfrak h})\otimes V)
   \end{equation}

\subsection{Use of mixed complex to compute the cohomology class responsible for obstacle}\label{sec:Detour}

The obstacle to covariance is $d_2\left({\bf 1}\in H^0(\mathfrak{g}/\mathfrak{h},{\cal O}\otimes H^2_{\rm BRST})\right)$,
using the spectral sequence of the
filtered complex (\ref{StraightForwardComplex}).
In Section \ref{FilteredComplex} we have shown that it is equivalent to the second page
of the ``effective'' complex $C^{\bullet}(\mathfrak{g},{\cal O})$, which can be presented as:
\begin{equation}
     C^n(\mathfrak{g}, {\cal O})
     \;=\;
     \bigoplus_{p+q=n}
     C^p\left(\,\mathfrak{g}/\mathfrak{h}\,,\, {\cal O}\otimes C^q(\mathfrak{h})\,\right)
     \label{StraightforwardComplex}\end{equation}
The spectral sequence of this ``effective'' complex is denoted $E^{p,q}_r$.

\includegraphics[scale=0.60000]{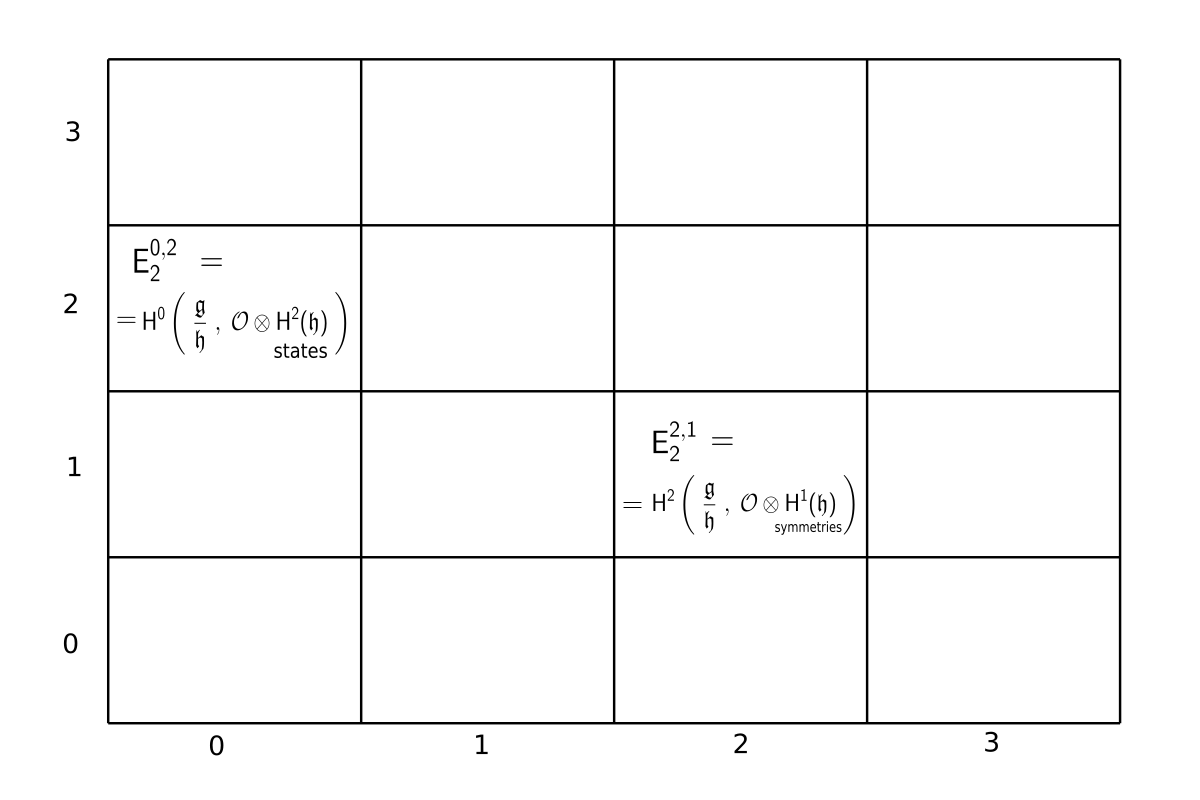}

On the other hand, the construction of
Section \ref{sec:MixedComplex}
gives us the ``mixed'' complex:
\begin{equation}
     \bigoplus_{p,q}
     C^p\left(\;\;
              {{\mathfrak g}/{\mathfrak h}}
              \;\;,\;\;
              C_q(\mathfrak{h})\otimes H^1(H\backslash \Pi TG)
              \;\right)
   \label{DetourComplex}\end{equation}                            
with the differential given by Eq. (\ref{MixedDifferential}).
The spectral sequence of this mixed complex will be denoted $\widetilde{E}_r^{p,-q}$ (where $q\geq 0$). We observe:
\begin{equation}
   \widetilde{E}_2^{2,-2} \simeq E_2^{2,1}
   \end{equation}
By construction, the obstacle to covariance is obtained as $d_2$ of an element of $E_2^{0,2}$.
But we can also obtain it as $\widetilde{d}_2$ of an element of $\widetilde{E}_2^{0,-1}$. The advantage is that  $\widetilde{E}_2^{0,-1}$ is finite-dimensional:
\begin{equation}
   \widetilde{E}_2^{0,-1} = H_1(\delta_{\mathfrak h})\otimes \mathfrak{sP} = \mbox{Hom}(\mathfrak{sP},\mathfrak{sP})
   \end{equation}
There is a canonical element, the identity map:
\begin{equation}
   {\bf 1}\in \mbox{Hom}(\mathfrak{sP},\mathfrak{sP})
   \end{equation}
We can obtain the obstacle cohomology class as $\widetilde{d}_2$ of this $\bf 1$.

\includegraphics[scale=0.75000]{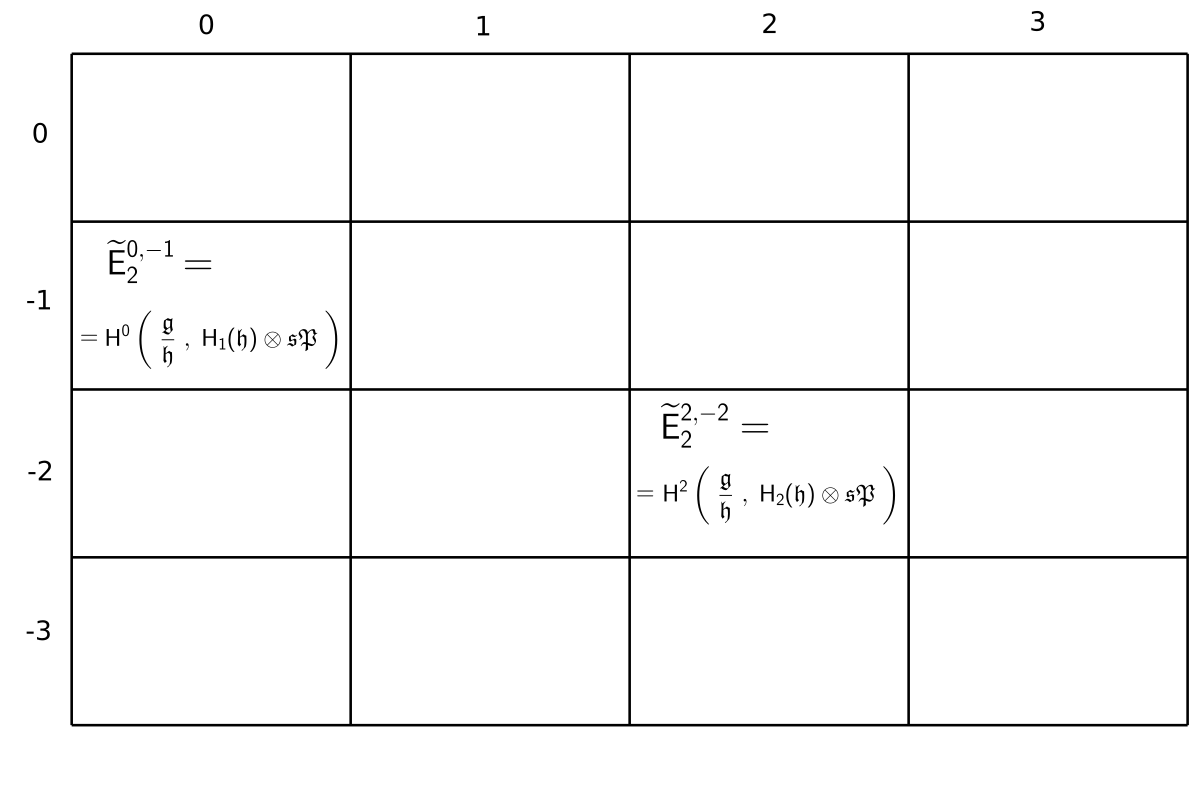}

This simplifies the computation, because all spaces involved are finite-dimensional.

\section{Computations}\label{Computations}

Here we will study the obstacle class:
\begin{align} \Psi \;\in\;
 & H^2\left(\;\mathfrak{g}/\mathfrak{h}\;,\;{\cal O}\otimes \mathfrak{sP}\;\right)
 \nonumber{} \\ \Psi \;=\;
 & \widetilde{d}_2\; ({\bf 1}\;:\; \mathfrak{sP}\rightarrow\mathfrak{sP})
 \nonumber{} \end{align}
The super-Poincare algebra $\mathfrak{sP}$ has an ideal $\mathfrak{susy}\subset\mathfrak{sP}$, and the factoralgebra
is the Lorentz algebra which we denote $\mathfrak{sP}/\mathfrak{susy} = \mathfrak{L}$. Therefore we can consider the projection:
\begin{equation}
   \Psi_{\mathfrak{L}}
   \in
   H^2\left(\;\mathfrak{g}/\mathfrak{h}\;,\;{\cal O}\otimes \mathfrak{L}\;\right)
   \;=\;
   H^2\left(\;\mathfrak{g}/\mathfrak{h}\;,\;{\cal O}\;\right)\otimes \mathfrak{L}
   \;=\;
   H^2\left(\;\mathfrak{g}/\mathfrak{h}\;,\;H_2(\mathfrak{h})\;\right)\otimes \mathfrak{L}
   \end{equation}
We will explicitly compute this projection here. 

\subsection{Preliminaries}\label{sec:ComputationPreliminaries}

\subsubsection{Lie superalgebra cohomology differential of $\mathfrak{susy} = \mathfrak{g}/\mathfrak{h}$}\label{sec:LieOperator}

\begin{equation}
   d_{\mathfrak{g}/\mathfrak{h}} =
   {1\over 2}c^m(\square^L_m + \square^R_m) + \gamma_L^{\alpha}\nabla^L_{\alpha} + \gamma_R^{\hat{\alpha}}\nabla^R_{\hat{\alpha}} -
   {1\over 2}((\gamma_L\Gamma^m\gamma_L) + (\gamma_R\Gamma^m\gamma_R)){\partial\over\partial c^m}
   \end{equation}

\subsubsection{Cohomological description of conserved currents}\label{sec:ConservedCurrentsInCohomology}

The identification of the ghost number one cohomology with global conserved currents has been
worked out in [\cite{Mikhailov:2012uh}]. It leads to the following identification:
\begin{align} {\bf 1}\;\in\;
 & H_1({\mathfrak h})\;\otimes\; \mathfrak{sP}
 \nonumber{} \\ {\bf 1}\;=\;
 & ([\square_m^L,\square_n^L] - [\square_m^R,\square_n^R])\otimes t_{[mn]} + (\square_m^L - \square_m^R)\otimes t_m + W_L^{\alpha}\otimes t_{\alpha}^L + W_R^{\hat{\alpha}}\otimes t_{\hat{\alpha}}^R
 \label{IdInGhostOne} \end{align}

\subsubsection{Obstacle in splitting out the ideal}\label{sec:ObstacleInSplittingIdeal}

The $\sigma$ of Eq. (\ref{DefSigma}) is given by:
\begin{align} \sigma \;=\;
 & \phantom{+\;} {1\over 8}c^mc^n([\square^L_m,\square^L_n] + [\square^R_m,\square^R_n]) \;+
 \nonumber{} \\  
 & +\; {1\over 2}\left([c\square^L,\gamma_L\nabla^L] + [c\square^R,\gamma_R\nabla^R]\right) \;+
 \nonumber{} \\  
 & +\; {1\over 4}((\gamma_L\Gamma^m\gamma_L) - (\gamma_R\Gamma^m\gamma_R))(\square^L_m - \square^R_m)
 \nonumber{} \end{align}
There is a cohomology class, preventing the ideal from splitting out:
\begin{equation}
   \sigma\;\mbox{mod}\;[\mathfrak{h},\mathfrak{h}] \;\in\;
   H^2(\,{\mathfrak g}/{\mathfrak h}\,,\,{\mathfrak h}/[{\mathfrak h},{\mathfrak h}])
   \end{equation}

\subsection{Computation of  $d_{\mathfrak{susy}}\delta_{\mathfrak h}^{-1}d_{\mathfrak{susy}}{\bf 1}$}\label{sec:BishopMove}

The most straightforward part is to compute the terms proportional to the Lorentz generators $t_{[mn]}\in\mathfrak{sP}$.
The coefficient of $\otimes t_{[mn]}$ in $\delta_{\mathfrak h}^{-1}d_{\mathfrak{susy}}{\bf 1}$ is:
\begin{align} \delta_{\mathfrak h}^{-1}d_{\mathfrak{susy}}{\bf 1}|_{t_{[mn]}}\;=\;
 & {1\over 2}(c\cdot(\square^L - \square^R))\wedge ([\square^L_m,\square^L_n] + [\square^R_m,\square^R_n])\;+
 \nonumber{} \\  
 & +\;2(\square^L_m - \square^R_m)\wedge (\gamma_L^{\alpha}[\nabla^L_{\alpha} ,\square^L_n] + \gamma_R^{\alpha}[\nabla^R_{\hat{\alpha}},\square^R_n])
 \nonumber{} \end{align}
Here $\ldots$ stands for the terms which contain only the generators of $\mathfrak{susy}\subset \mathfrak{sP}$, but not $t_{[mn]}$.
Since we are only computing the projection to $\mathfrak{L}$, we do not need those terms for our computation here.
We complete the ``bishop move'' by applying $d_{\mathfrak{susy}}$. The coefficient of $\otimes t_{[mn]}$ is:
\begin{align} d_{\mathfrak{susy}}\delta_{\mathfrak h}^{-1}d_{\mathfrak{susy}}{\bf 1}|_{t_{[mn]}}\;=\;
 & \phantom{+\;}
        {1\over 4}
        c^p c^q ([\square^L_p,\square^L_q] - [\square^R_p,\square^R_q])
        \wedge
        ([\square^L_m,\square^L_n] + [\square^R_m,\square^R_n])\;-
 \nonumber{} \\  
 & -\;{1\over 4}(c\square^L - c\square^R)\wedge
        ([c\square^L,[\square^L_m,\square_n^L]] + [c\square^R,[\square^R_m,\square^R_n]])\;+
 \nonumber{} \\  
 & +\;([c \square^L,\square^L_m] - [c \square^R,\square^R_m])
        \wedge
        ([\gamma_L\nabla^L ,\square^L_n] + [\gamma_R\nabla^R,\square^R_n])
 \nonumber{} \\  
 & +\;(\square_m^L - \square_m^R)\wedge ([c\square^L,[\gamma_L\nabla^L,\square^L_n]] + [c\square^R,[\gamma_R\nabla^R,\square^R_n]])
 \nonumber{} \\  
 & -\;{1\over 2}(c\square^L - c\square^R)
        \wedge
        \left([\gamma_L\nabla^L,[\square^L_m,\square^L_n]]
         +
         [\gamma_R\nabla^R,[\square^R_m,\square^R_n]]\right)
 \nonumber{} \\  
 & +\;{1\over 2}[(\gamma_L \nabla^L + \gamma_R\nabla^R),(c\square^L - c\square^R)]
           \wedge
           ([\square^L_m,\square^L_n] + [\square^R_m,\square^R_n])\;+
 \nonumber{} \\  
 & +\;2([\gamma_L\nabla^L ,\square^L_m] -[\gamma_R\nabla^R,\square^R_m])
            \wedge
            ([\gamma_L\nabla^L ,\square^L_n] + [\gamma_R\nabla^R,\square^R_n])\;+
 \nonumber{} \\  
 & +\;(\square^L_m - \square^R_m)\wedge
           ((\gamma_L\Gamma^k\gamma_L)[\square_k^L,\square_n^L] + (\gamma_R\Gamma^k\gamma_R)[\square_k^R,\square_n^R])
 \nonumber{} \\  
 & -{1\over 4}\;((\gamma_L\Gamma^k\gamma_L) + (\gamma_R\Gamma^k\gamma_R))(\square^L_k - \square^R_k)\wedge ([\square^L_m,\square^L_n] + [\square^R_m,\square^R_n])
 \nonumber{} \end{align}  
Finally, the whole $\widetilde{d}_2$ is:
\begin{align}  
 & \left(-\sigma\wedge + d_{\mathfrak{susy}}\delta_{\mathfrak h}^{-1}d_{\mathfrak{susy}}\right){\bf 1}|_{t_{[mn]}}\;=\;
 \nonumber{} \\ =\;
 & \phantom{+\;}
           {1\over 4}
           ([c\square^L,c\square^L] - [c\square^R,c\square^R])
           \wedge
           ([\square^L_m,\square^L_n] + [\square^R_m,\square^R_n])\;-
 \nonumber{} \\  
 & -\;
        {1\over 8}
        ([c\square^L,c\square^L] + [c\square^R,c\square^R])
        \wedge
        ([\square^L_m,\square^L_n] - [\square^R_m,\square^R_n])\;-
 \nonumber{} \\  
 & - \;{1\over 4}(c\square^L - c\square^R)\wedge
        ([c\cdot\square^L,[\square_m^L,\square_n^L]] + [c\cdot\square^R,[\square_m^R,\square_n^R]])\;+
 \nonumber{} \\  
 & +\;([c \square^L,\square^L_m] - [c \square^R,\square^R_m])
        \wedge
        ([\gamma_L\nabla^L ,\square^L_n] + [\gamma_R\nabla^R,\square^R_n])
 \nonumber{} \\  
 & +\;(\square_m^L - \square_m^R)\wedge ([c\square^L,[\gamma_L\nabla^L,\square^L_n]] + [c\square^R,[\gamma_R\nabla^R,\square^R_n]])
 \nonumber{} \\  
 & -\;{1\over 2}(c\square^L - c\square^R)
        \wedge
        \left([\gamma_L\nabla^L,[\square^L_m,\square^L_n]]
         +
         [\gamma_R\nabla^R,[\square^R_m,\square^R_n]]\right)
 \nonumber{} \\  
 & +\;{1\over 2}[(\gamma_L \nabla^L + \gamma_R\nabla^R),(c\square^L - c\square^R)]
           \wedge
           ([\square^L_m,\square^L_n] + [\square^R_m,\square^R_n])\;+
 \nonumber{} \\  
 & -\; {1\over 2}([c\square^L,\gamma_L\nabla^L] + [c\square^R,\gamma_R\nabla^R])\wedge ([\square_m^L,\square_n^L] - [\square_m^R,\square_n^R]) \;+
 \nonumber{} \\  
 & +\;4[\gamma_L\nabla^L ,\square^L_m]
            \wedge
            [\gamma_R\nabla^R ,\square^R_n]
            \;+
 \nonumber{} \\  
 & +\;(\square^L_m - \square^R_m)\wedge
           ((\gamma_L\Gamma^k\gamma_L)[\square_k^L,\square_n^L] + (\gamma_R\Gamma^k\gamma_R)[\square_k^R,\square_n^R])
 \nonumber{} \\  
 & -{1\over 4}\;((\gamma_L\Gamma^k\gamma_L) + (\gamma_R\Gamma^k\gamma_R))(\square^L_k - \square^R_k)\wedge ([\square^L_m,\square^L_n] + [\square^R_m,\square^R_n])
 \nonumber{} \end{align}
This can be simplified, modulo $\delta$-exact terms, as follows:
\begin{align}  
 & {1\over 2}[c\square^L,c\square^L]\wedge [\square^R_m, \square^R_n] - {1\over 2}[c\square^R,c\square^R]\wedge [\square^L_m, \square^L_n]
 \nonumber{} \\  
 & +\;
        2[c\square^L,\square^L_m]\wedge [\gamma_R\nabla^R,\square^R_n]
        - 2[c\square^R,\square^R_m]\wedge [\gamma_L\nabla^L,\square^L_n]
 \nonumber{} \\  
 & +\;
        [c\square^L,\gamma_L\nabla^L]\wedge [\square^R_m,\square^R_n]
        - [c\square^R,\gamma_R\nabla^R]\wedge [\square^L_m,\square^L_n]
 \nonumber{} \\  
 & +\;4[\gamma_L\nabla^L ,\square^L_m]
            \wedge
            [\gamma_R\nabla^R ,\square^R_n]
            \;+
 \nonumber{} \\  
 & +\;{1\over 2}((\gamma_L\Gamma^k\gamma_L)-(\gamma_R\Gamma^k\gamma_R))
           \left(
                 (\square^L-\square^R)_{[k}\wedge [\square^L_m,\square^L_{n]}]
                 -
                 (\square^L-\square^R)_{[k}\wedge [\square^R_m,\square^R_{n]}]
                 \right)
 \nonumber{} \end{align}
This can be rewritten in terms of the SUGRA fields;
schematically:
\begin{align}  
 & c^pc^q R_{[pq][mn]}
 \nonumber{} \\  
 & + c^p((\partial_{[p}\psi^L_{m]})\Gamma_n\gamma_L - (\partial_{[p}\psi^R_{m]})\Gamma_n\gamma_R - (m\leftrightarrow n))
        + (\gamma_L\hat{c}\partial_{[m}\psi^L_{n]}) - (\gamma_R\hat{c}\partial_{[m}\psi^R_{n]})
 \nonumber{} \\  
 & + (\gamma_L \Gamma_{[m} \hat{F} \Gamma_{n]} \gamma_R) + ((\gamma_L\Gamma^k\gamma_L) - (\gamma_R\Gamma^k\gamma_R))H_{kmn}
 \nonumber{} \end{align}
where $R_{[pq][mn]}$ is the Riemann tensor,
$\widehat{F}$  the RR bispinor, $H_{kmn}$  the NSNS field strength, and $\psi_m$  gravitinos.

\subsection{The case of super-Maxwell theory}\label{sec:ComputationForMaxwell}

There is a similar computation for super-Maxwell theory. Instead of $\mathfrak{h}\subset\mathfrak{g}$
we have $I\subset L$. The space of classical solutions is $H^1_{\rm BRST}$
(while for linearized SUGRA it was $H^2_{\rm BRST}$). We start with the identity map:
\begin{equation}
   {\bf 1}\in E_2^{0,1} = ({\cal O}\otimes H^1_{\rm BRST})^{L/I}
   \end{equation}
The obstacle is $d_2 {\bf 1}$, it lives in $E_2^{2,0}$.

\includegraphics[scale=0.75000]{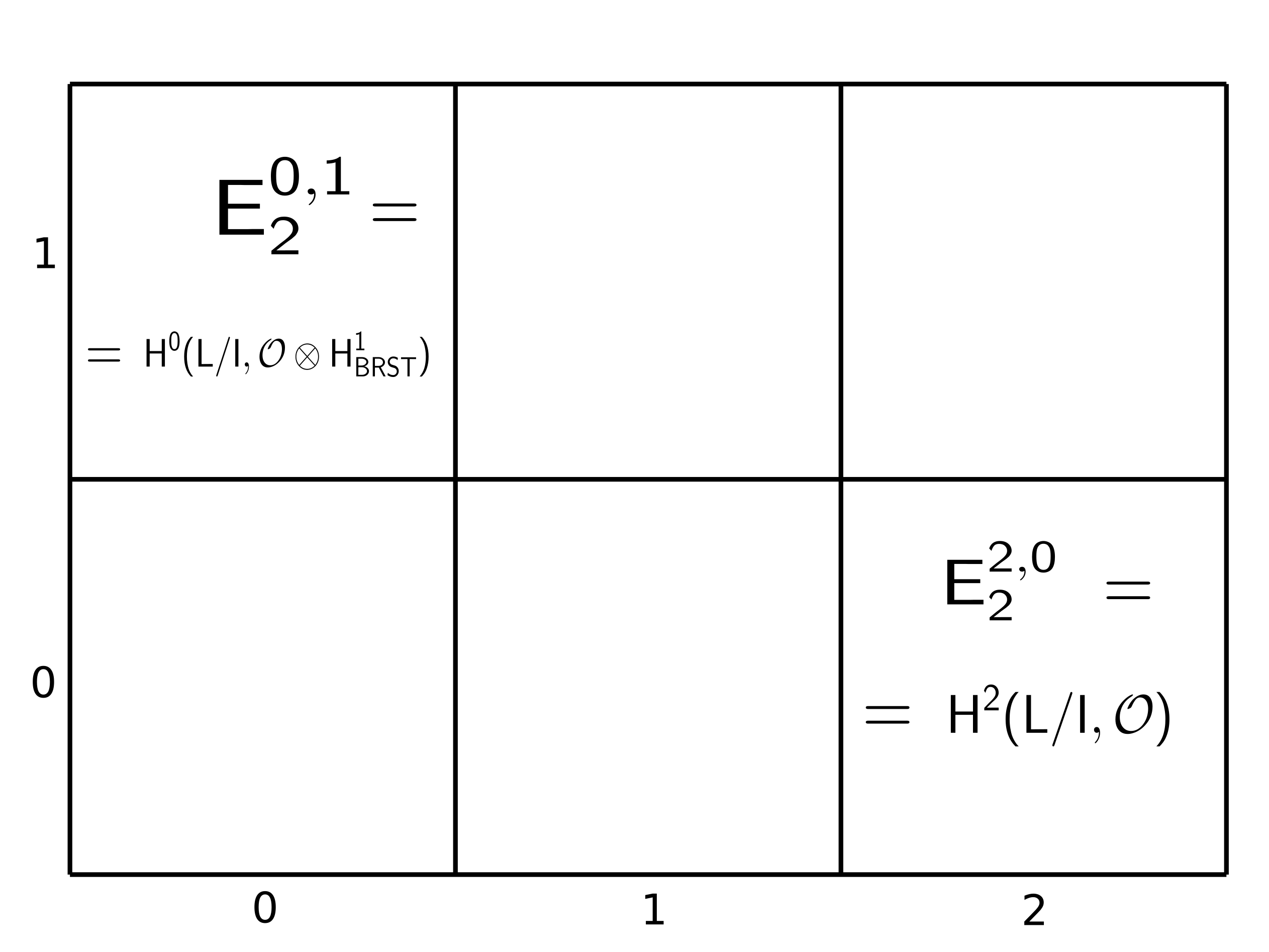}

We can compute the obstacle cohomology class using the mixed complex,
since $E_2^{2,0}=\widetilde{E}_2^{2,-1}$.
We need:
\begin{equation}
   \widetilde{d}_2\;:\;\widetilde{E}_2^{0,0} \rightarrow \widetilde{E}_2^{2,-1}
   \end{equation}
Since $\widetilde{E}_2^{0,0}$ is one-dimensional, $d \delta^{-1} d$ is zero and the obstacle is just $\sigma$.

\includegraphics[scale=0.75000]{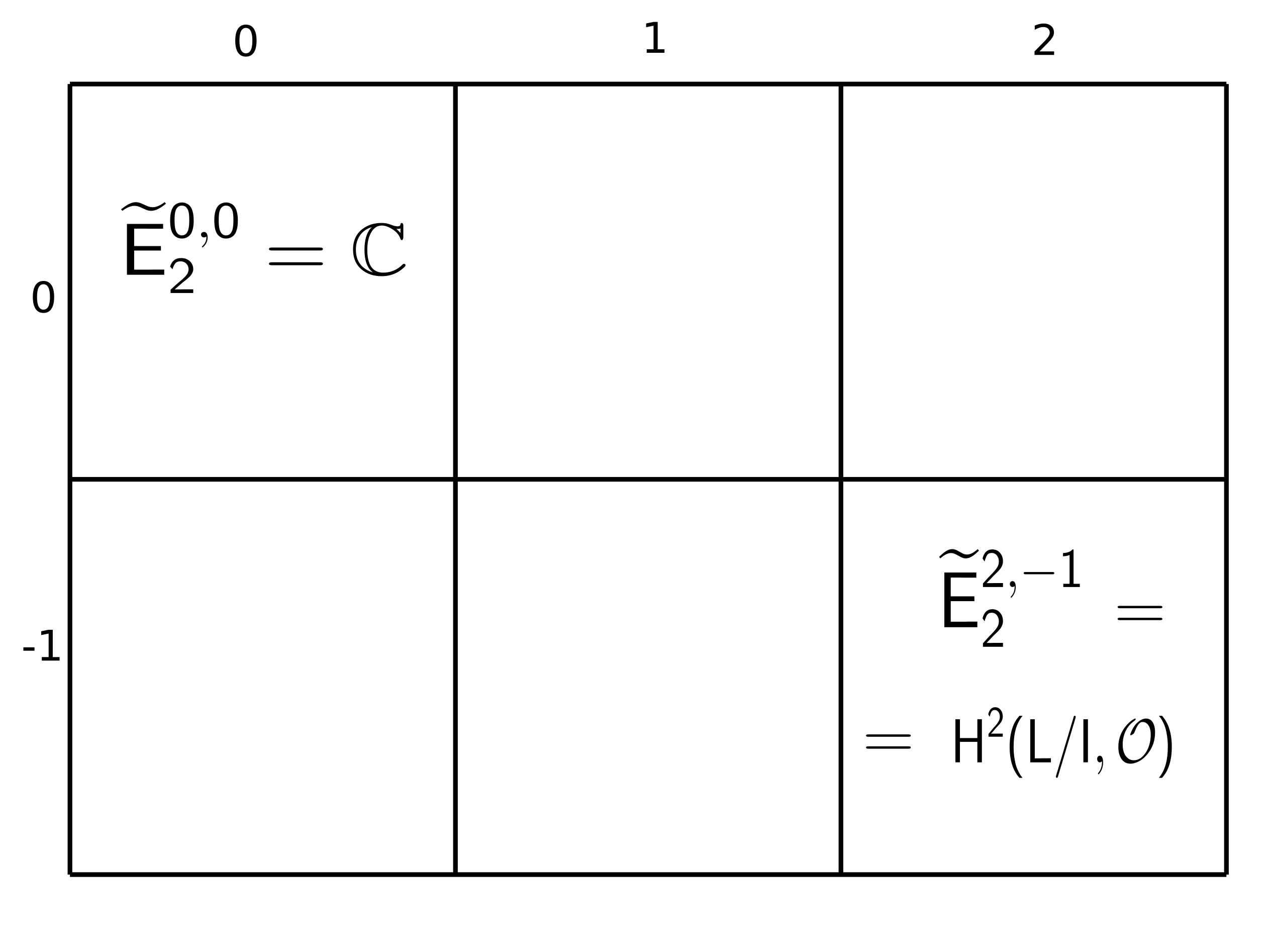}

\section{Acknowledgments}\label{Acknowledments}

This work was supported in part by  ICTP-SAIFR FAPESP  grant 2019/21281-4,
and in part by CNPq grant ``Produtividade em Pesquisa'' 307191/2022-2.

\def\cprime{$'$} \def\cprime{$'$}
\providecommand{\href}[2]{#2}\begingroup\raggedright\endgroup
\end{document}